\documentclass[prx,twocolumn,amsart,superscriptaddress]{revtex4-1}
\usepackage{graphicx}
\usepackage{amssymb}
\usepackage{amsmath}
\usepackage{color}
\usepackage{verbatim}
\usepackage{mathrsfs}
\usepackage{slashed}
\usepackage{esint}
\usepackage{hyperref}
\usepackage{tabularx}
\usepackage{epstopdf}

\newcommand{\matr}[1]{\mathsf{#1}}

\newtheorem{theorem}{Theorem}

\begin{document}

\title{Strictly local tensor networks for short-range topological insulators}

\author{Shaoyu Yin}
\affiliation{School of Physics \& Astronomy, University of Birmingham, Edgbaston, Birmingham, B15 2TT, United Kingdom}
\affiliation{Institute for Theoretical Physics and Cosmology, Zhejiang University of Technology, Hangzhou 310023, China}
\author{Nigel R. Cooper}
\affiliation{T.C.M. Group, Cavendish Laboratory, University of Cambridge, J.J. Thomson Avenue, Cambridge, CB3 0HE, United Kingdom}
\author{Benjamin B\'eri}
\affiliation{School of Physics \& Astronomy, University of Birmingham, Edgbaston, Birmingham, B15 2TT, United Kingdom}
\affiliation{T.C.M. Group, Cavendish Laboratory, University of Cambridge, J.J. Thomson Avenue, Cambridge, CB3 0HE, United Kingdom}
\affiliation{DAMTP, University of Cambridge, Wilberforce Road, Cambridge, CB3 0WA, United Kingdom}

\begin{abstract}
Despite the success in describing a range of quantum many-body states using tensor networks, there is a no-go theorem that rules out strictly local tensor networks as topologically nontrivial groundstates of gapped parent Hamiltonians with short-range (including exponentially decaying) couplings. In this work, we show that for free fermions, strictly local tensor networks may describe nonzero temperature averages with respect to gapped Hamiltonians with exponentially decaying couplings. Parent Hamiltonians in this sense may be constructed for any dimensionality and without any obstructions due to their topology. Conversely, we also show that thermal averages with respect to gapped, strictly short-range free-fermion Hamiltonians can be calculated by tensor networks whose links decay exponentially with distance. We also describe a truncation-reconstruction scheme for such tensor networks that leads to a controlled approximation of exact averages in terms of a sequence of related thermal averages. We illustrate our scheme on the two-dimensional Haldane honeycomb model considering both topological and nontopological phases.
\end{abstract}

\date{December 2018}
\maketitle

\vspace*{-4em}

\section{Introduction}

It is highly nontrivial to obtain the eigenstates in general quantum many-body problems, even for the ground state. Among the possible approaches, tensor networks \cite{TensorNetwork1,TensorNetwork2,ground} have established themselves as promising generalizations of matrix-product-state methods \cite{White:DMRG,Schollwoeck:2011,TEBD} to systems beyond one spatial dimension. Conceptual and methodological aspects of tensor networks are rapidly developing, with scope including not only condensed matter systems, but also fields such as quantum chemistry \cite{Nakatani:2013}, holographic dualities\cite{holographic}, etc. The efficiency and versatility of tensor networks is due to their capability of encoding correlations and entanglement in quantum states such that they effectively capture the relevant Hilbert space based on the area law of entanglement entropy\cite{area-law}. Thus tensor networks can be viewed as a natural language for quantum problems where area laws are relevant \cite{Orus:2014}.

Tensor networks are parameterized by tensors on discrete sites, and links between them describing the contraction of indices of tensors at the link ends. For a tensor network description to be efficient, the degrees of freedom (DOF) in the tensor variables should not grow exponentially with system size. This, in particular, rules out tensor networks with long-range link structure, since each additional link to a tensor implies an additional tensor index.
For gapped systems, the area law, as well as the exponential decay of correlations \cite{Hastings:2004,Kitaev:2006} suggest some local structure for ground states. One may thus wonder whether a description is possible using \textit{local} tensor networks, where links between sites are unimportant, or even absent, beyond a certain finite range.

Such local tensor networks, may however be at odds with topologically nontrivial states, as these display characteristic features largely independent of local details. Indeed, even though a range of time-reversal symmetric topological states (e.g. string-net condensates \cite{stringnet}) are known to admit local tensor network representations, obstacles exist \cite{Wahl:2013,Dubail:2015} in the local tensor network description of chiral topological states (e.g. quantum Hall states or chiral $p$-wave superconductors), and more generally for any topologically nontrivial free-fermion state beyond one dimension \cite{Read:2017}. Recent no-go theorems \cite{Wahl:2013,Dubail:2015,Read:2017} in fact  show that local tensor networks for such topological states can only be the ground states of parent Hamiltonians
 \cite{parent,Wahl:2013,Yang:2015} that are either gapless or have power-law-decaying couplings.
 (Useful, albeit nonlocal, tensor networks for such states with gapped short-range parent Hamiltonians \cite{Iblisdir:2007,Feiguin:2008,Changlani:2009,Motruk:2016}, 
 as well as numerical studies using strictly local tensor networks that account for spurious long-range correlations \cite{Chen:2018}, do however exist.)

An alternative approach to the tensor network description of topological phases was suggested in Ref.~\onlinecite{Beri:2011}. Instead of constructing ground \textit{states}, %
it was shown that there exist simple local tensor network representations of groundstate {\it expectation values} for a range of topological states, including chiral and strongly correlated cases.

The tensor networks of Ref.~\onlinecite{Beri:2011} are \textit{``exponentially local"}: the importance of their links decays exponentially with distance. This is distinct from the \textit{``strictly local"} tensor networks in the no-go theorems \cite{Wahl:2013,Dubail:2015,Read:2017} mentioned earlier; there links are entirely absent beyond a certain finite distance.
Such strictly local tensor networks also naturally arise in applying the approach of Ref.~\onlinecite{Beri:2011} in practice, since keeping tensor sizes from scaling exponentially with system size requires truncation.

Motivated by these observations, in this work we ask where the tensor networks for expectation values \cite{Beri:2011} lead us if we insist that the networks therein be strictly local. Focusing on free fermions as in Refs.~\cite{Wahl:2013,Dubail:2015,Read:2017}, we consider two complementary perspectives:

(i) In the spirit of the no-go theorems \cite{Wahl:2013,Dubail:2015,Read:2017}, we take a tensor network as the starting point, and ask about the parent Hamiltonian. We show that a parent Hamiltonian may indeed be defined; with respect to this the tensor network describes \textit{nonzero temperature} averages. Remarkably, the parent Hamiltonians may be both gapped and have exponentially short-range couplings with no obstructions imposed by their topology.

(ii) Conversely, in the spirit of Ref.~\onlinecite{Beri:2011}, we take a gapped strictly short-range Hamiltonian $\hat{H}$ as a starting point. We show that, under a suitable truncation-reconstruction procedure, an exact exponentially local tensor network for groundstate averages is approximated by a strictly local network for nonzero temperature averages with respect to a parent Hamiltonian.
This process is expected to be exponentially controlled: the elimination of exponentially small couplings should only have small quantitative effects.
Reducing the degree of truncation (i.e., increasing the  range of permitted network links) corresponds to reducing the fictitious temperature and approaching $\hat{H}$ with the parent Hamiltonian.
Throughout this process, the parent Hamiltonian remains gapped; for gapped systems, this truncation-reconstruction procedure thus preserves topological features and connects smoothly to the ground state without significantly affecting physical properties. The same considerations also apply for nonzero temperature averages \mbox{with respect to $\hat{H}$.}

Our results are based on a generalization of the approach of Ref.~\onlinecite{Beri:2011} to describe the expectation values not just in the ground state but at general nonzero temperatures. We therefore start the presentation, in Sec.~\ref{sec:GTN}, by explaining the basis of constructing tensor networks for thermal averages with respect to  gapped free fermion systems. In Sec.~\ref{sec:theoremHpar}, we show how suitable mappings between Hamiltonians and such tensor networks
relate strictly local to exponentially local objects, including strictly local tensor networks to gapped, exponentially short-range, parent Hamiltonians for thermal averages [(i) above].
In Sec.~\ref{sec:trunc-rec-gen}, we discuss how parent Hamiltonians for thermal averages can be reconstructed after truncating an exponentially local tensor network [(ii) above]. We numerically illustrate this in Sec.~\ref{sec:numerics} taking the two-dimensional (2D) Haldane model \cite{Haldane:1988} as an example. We close in Sec.~\ref{sec:summary} with a summary and discussion. The Appendices contain further details on the Grassmann expression underlying our tensor network averages (App.~\ref{app:grassmann}), topological aspects of thermal states (App.~\ref{app:rhotop}), and the outline of the proof of the theorem invoked in Sec.~\ref{sec:theoremHpar} establishing the strictly local to exponentially local correspondence (App.~\ref{app:bcb}).

\section{Grassmann tensor networks and free fermion thermal averages}
\label{sec:GTN}

We consider a free fermion system with quadratic Hamiltonian $\hat{H} = \sum_{i,j} \hat{a}_i^\dag H_{ij} \hat{a}_j$, where $\hat{a}_i^{(\dag)}$ are fermionic annihilation (creation) operators on a set of $n$ lattice sites labelled by $i=1,\ldots n$. We denote the eigenstates of the single-particle Hamiltonian matrix  $\matr{H}$ (with components  $H_{ij}$ in the basis of lattice sites) by $|\psi_k\rangle$ and the corresponding energies (measured from the chemical potential) by $\epsilon_k$. The many-body ground state $|\Psi\rangle$ is the Slater determinant of all occupied ($\epsilon_k\le0$) eigenstates. We will be considering insulators: particle number conserving gapped systems such that $\epsilon_k\neq0$ for all system sizes, i.e., the chemical potential lies within the gap of the system. (Our considerations can be also straightforwardly extended to gapped superconductors.) It will be useful to introduce the single-particle spectral projector $\matr{P} =\sum_{\epsilon_k<0} |\psi_k\rangle\langle\psi_k| $ onto the single-particle states occupied in the ground state. The matrix representation of this projector in the lattice basis describes two point groundstate correlations of the lattice operators, $P_{ij} =\langle\Psi|\hat a_j^\dagger\hat a_i|\Psi\rangle$. It will also be useful to introduce the so-called flat band Hamiltonian $\matr{h} =1-2\matr{P}=\text{sgn}\,{\matr{H}}$ \cite{Kitaev:2006},
a matrix obtained from the single-particle Hamiltonian $\matr{H}$ by flattening its spectrum to $-1$ ($+1$) for occupied (unoccupied) states, but keeping intact its gapped nature and eigenvectors, and hence its topological features. 
The tensor networks of Ref.~\cite{Beri:2011} for groundstate expectation values  are based on $\matr{h}$.

For  nonzero temperatures $\beta^{-1}$, the system can be described by the normalized density matrix \mbox{$\hat\rho=\frac{1}{Z}\mathrm e^{-\beta\hat H}=\frac{1}{Z}\mathrm e^{-\sum_k\beta\epsilon_k\hat a_k^\dagger\hat a_k}$} with partition function \mbox{$Z=\prod_k(1+\mathrm e^{-\beta\epsilon_k})$} and annihilation (creation) operators $\hat a_k^{(\dagger)}$ of eigenstates $|\psi_k\rangle$. Using the same Grassmann integral approach \cite{Bravyi:2005} as in Ref.~\onlinecite{Beri:2011}, we next show that the expressions for nonzero temperature thermal expectation values have the same structure, and thus relation to Grassmann tensor networks \cite{Gu:2010}, as those in Ref.~\onlinecite{Beri:2011} for groundstate averages.

The thermal average $\langle \hat{X}(\hat a,\hat a^\dagger)\rangle\!=\!\mathrm{Tr}[\hat{X}(\hat a,\hat a^\dagger)\hat\rho]$ of an operator $\hat{X}(\hat a,\hat a^\dagger)$
can be conveniently expressed as an integral over a set of Grassmann variables. Following Bravyi \cite{Bravyi:2005} and Ref.~\onlinecite{Beri:2011}, the Grassmann mapping substitutes
$\frac{\theta_j}{\sqrt{2}}$ for $\hat a_j$, $\frac{\bar\theta_j}{\sqrt{2}}$ for $\hat a_j^\dagger$, and $\frac{\bar\theta_j\theta_j+1}{2}$ for $\hat a_j^\dagger\hat a_j$, where $\theta_j$ and $\bar\theta_j$ are  Grassmann variables. (We assume that operators on the same site appear in the order $\hat a_j^\dagger\hat a_j$, without loss of generality.) This maps $\hat{X}(\hat a,\hat a^\dagger)$ to some function
$\tilde X(\theta,\bar{\theta})$.  As shown in Appendix~\ref{app:grassmann},
\begin{equation}\label{Grassmann}
\langle \hat{X}(\hat a,\hat a^\dagger)\rangle=N\int(\prod_{j=1}^n\mathrm d\theta_j\mathrm d\bar\theta_j)\tilde X(\theta,\bar{\theta})\mathrm e^{-\bar{\boldsymbol\theta}^\top\matr{h}_T\boldsymbol\theta},
\end{equation}
where
\begin{equation}\label{norm}
N^{-1}=\int(\prod_{j=1}^n\mathrm d\theta_j\mathrm d\bar\theta_j)\mathrm e^{-\bar{\boldsymbol\theta}^\top\matr{h}_T\boldsymbol\theta}=\det(-\matr{h}_T)
\end{equation}
is an $X$ independent normalization factor and
\begin{equation}\label{hH}
\matr{h}_T=\coth\left(\frac{\beta}{2}\matr{H}\right).
\end{equation}
Importantly, Eqs.~\eqref{Grassmann} and \eqref{norm} are almost identical to the Grassmann integrals that form the starting point of the analysis in Ref.~\onlinecite{Beri:2011};
the only difference is the appearance of $\matr{h}_T$ instead of $\matr{h}$. [In the zero temperature limit, $\beta\rightarrow\infty$, $\epsilon_k\neq0$ and $\coth(\pm \infty)=\pm1$ ensures $\matr{h}_T\rightarrow \matr{h}$.] It is based on these expressions that Ref.~\onlinecite{Beri:2011} develops a Grassmann tensor network \cite{Gu:2010} description and the same approach thus applies to the nonzero temperature case that we consider here. The key step involves \cite{Beri:2011} the Grassmann Leibniz rule \cite{Zinn-Justin:2002} to change the integral over Grassmann variables on sites to variables on links, with a nonzero matrix element $(\matr{h}_T)_{jl}$ corresponding to there being a link between sites $j$ and $l$. For our considerations below, the transformation into a link-variable structure is not necessary, and thus it will be Eq.~\eqref{Grassmann} that will form the basis of our subsequent analysis. We will, however, use that it is $\matr{h}_T$ that determines the link structure of the tensor network we could convert to, and thus we shall henceforth refer to $\matr{h}_T$ as the link matrix. (For completeness, in Appendix~\ref{app:GTNrel} we briefly summarize the explicit relation between our expressions and the Grassmann tensor network contractions of Ref.~\onlinecite{Gu:2010}.)

\section{Strictly local tensor networks and their parent Hamiltonians}
\label{sec:theoremHpar}

Having identified $\matr{h}_T$ as the key object encoding the tensor network connectivity properties, we can now turn to discussing how the locality properties of corresponding  Hamiltonian -- tensor network pairs are related. The basis of this relation is the following theorem on ``strictly short-range" matrices, i.e., matrices $\matr{M}$ with indices $i$, $j$
labeling the sites of a finite lattice, and for which there is a range $R^*>0$ much smaller than the linear size of the lattice such that $M_{ij}=0$ if $R_{ij}>R^*$, where $R_{ij}$ denotes the distance between sites $i$ and $j$:
\begin{theorem}
\label{th:1}
Let $\matr{M}$ be a Hermitian strictly short-range matrix with range $R^*$ and eigenvalues within $D\!=\!D_-\!\cup\!D_+$ where \mbox{$D_-\!=\![-\eta_2,-\eta_1]$, $D_+\!=\![\eta_1,\eta_2]$} with $0<\eta_1<\eta_2$. Let $f$ be such that on $D_-$ ($D_+$) it is the restriction of a function $f_1$ ($f_2$) analytic in a neighborhood of $D_-$ ($D_+$) in the complex plane.
Then, there exist  $C>0$ and $\chi>1$ such that
\begin{equation}
|f(\matr{M})_{ij}|\leq C \chi^{-R_{ij}/R^*}.
\end{equation}
\end{theorem}
This result is based on the work of Benzi {\it et al.} \cite{Benzi:2012}, who proved a variant of it [for $f_1$ ($f_2$) analytic on the left (right) half of the complex plane] building on approximation theorems of Chui and Hasson~\cite{Chui:1983} and Bernstein \cite{Bernstein}. Our formulation of the theorem differs from that of Ref.~\onlinecite{Benzi:2012} only in qualifying the domain of analyticity of $f_{1,2}$ and is thus proven following nearly the same steps. We outline the proof in Appendix \ref{app:bcb}.

The utility of the Theorem \ref{th:1} for our purposes lies in taking $\matr{M}$ as the strictly short-range object that we wish to relate to its counterpart. Taking $\matr{M}=\matr{H}$, the gapped strictly short-range single-particle Hamiltonian, the exponentially local nature of the tensor network for groundstate averages \cite{Beri:2011} follows if we take $f(x)=\text{sgn}(x)$, that is $f(\matr{H})=\matr{h}$, the flat band Hamiltonian, which, as we noted in Sec.~\ref{sec:GTN}, is the groundstate link matrix. (The exponential decay of $\matr{h}$ was also noted in Ref.~\onlinecite{Kitaev:2006} in a form of an approximation theorem; see also Ref.~\onlinecite{Ringel:2011}.)

If we again take $\matr{M}=\matr{H}$ but now we use \mbox{$f(x)=\coth(\beta x/2)$}, that is $f(\matr{H})=\matr{h}_T$, we find that the thermal averages remain described by exponentially local tensor networks even for nonzero temperatures.

A particularly interesting implication of Theorem \ref{th:1} arises when one applies it in reverse: take a tensor network \textit{defined} via its link matrix $\matr{g}$. Requiring the  spectrum of $\matr{g}$ to lie within the range of $\coth$ ensures that
\begin{equation}\label{GTN-tostart}
\langle \hat{X}(\hat a,\hat a^\dagger)\rangle=N\int(\prod_{j=1}^n\mathrm d\theta_j\mathrm d\bar\theta_j)\tilde X(\theta,\bar{\theta})\mathrm e^{-\bar{\boldsymbol\theta}^\top\matr{g}\boldsymbol\theta},
\end{equation}
with
\begin{equation}
N^{-1}=\int(\prod_{j=1}^n\mathrm d\theta_j\mathrm d\bar\theta_j)\mathrm e^{-\bar{\boldsymbol\theta}^\top\matr{g}\boldsymbol\theta}=\det(-\matr{g}),
\end{equation}
corresponds to the thermal average with respect to a parent  Hamiltonian $\matr{H}_\text{par}$ defined by
\begin{equation}\label{eq:Hpar}
\beta_\text{par} \matr{H}_\text{par}=2\,\text{arccoth}(\matr{g}).
\end{equation}
We note that this procedure defines $\matr{H}_\text{par}$, and thus also the temperature $\beta_\text{par}^{-1}$, only up to an overall energy scale. However, this already determines the decay properties of the couplings and the presence or absence of a spectral gap in $\matr{H}_\text{par}$, which are the features we are interested in.

Suppose now that $\matr{g}$ defines a strictly local tensor network, and that the spectrum of $\matr{g}$ is contained within the union of finite intervals
$[-g_\text{max},-g_\text{min}]\cup[g_\text{min},g_\text{max}]$, $1\!<\!g_\text{min},g_\text{max}\!<\!\infty$ such that the number of eigenvalues contained in each of the intervals increases linearly with the number of sites $n$. (That is, viewed as a single-particle Hamiltonian, $\matr{g}$ is a strictly short-range finite bandwidth insulator with spectrum in the analytic domain of arccoth.) Then, by Theorem~\ref{th:1} and Eq.~\eqref{eq:Hpar}, $\matr{H}_\text{par}$ will be an insulator with exponentially decaying couplings. Furthermore, the eigenvectors of $\matr{g}$ and $\matr{H}_\text{par}$ are the same; in particular the eigenvectors of $\matr{g}$ corresponding to eigenvalues $G_j$ of a given sign are eigenvectors of $\matr{H}_\text{par}$ with eigenvalues of the same sign. Therefore, if the $G_j<0$ eigenvectors of $\matr{g}$ have topological features, including being those of a chiral topological state with nonzero Chern number, then $\matr{H}_\text{par}$ will define a ground state, and thermal density matrix, with the same topological features. (The density matrix topology follows from the same eigenvectors \cite{rhotop1,rhotop2}, as we review in Appendix~\ref{app:rhotop}.)
We thus find that thermal expectation values of gapped parent Hamiltonians with exponentially decaying couplings can be represented by strictly local tensor networks, without any restrictions imposed by topology.

It is useful to compare this finding to the no-go theorems of Refs.~\onlinecite{Dubail:2015,Read:2017}: these rule out the existence of
gapped parent Hamiltonians with short-range  (including exponentially decaying) couplings for a large class of strictly local tensor networks  encoding topologically nontrivial \textit{ground states}. Our construction, in contrast, works with different settings: firstly we consider expectation values instead of a direct representation of states. Secondly, at least for topologically nontrivial insulators $\matr{H}_\text{par}$, the strictly local tensor networks in our approach always correspond to thermal averages at strictly positive temperature. This is not a mere artefact of using Eq.~\eqref{eq:Hpar} to define $\matr{H}_\text{par}$: for any $\matr{H}_\text{par}$, zero temperature averages correspond to $\matr{g}=\text{sgn}\, \matr{H}_\text{par}$, which cannot be both topologically nontrivial and strictly short-range \cite{Chen:2013,Read:2017}.

\section{Truncation and Reconstruction}
\label{sec:trunc-rec-gen}

Starting from an exponentially local tensor network, a strictly local tensor network naturally arises if one wishes to perform a numerical computation:
In a practical calculation, to have a numerically tractable number of DOF, one has to truncate some of the longer links, be the tensor network constructed from the flat band Hamiltonian $\matr{h}$ in the zero temperature case \cite{Beri:2011}, or from $\matr{h}_T$   at nonzero temperatures. We shall refer to the resulting truncated link matrix as $\matr{h}^{\rm t}$.
If the goal is simply to obtain a controlled approximation for the purposes of numerical results, one may obtain $\matr{h}^{\rm t}$ by directly discarding links beyond a certain cutoff length; the error will be controlled exponentially due to the exponentially local nature of the exact network.
However, more care should be taken if one wants to associate a valid free-fermion state to the network after truncation, because merely truncating $\matr{h}$ or $\matr{h}_T$ may place some of the eigenvalues $\eta^{\rm t}_k$ of $\matr{h}^{\rm t}$ in the interval  $(-1,1)$, thus placing the spectrum of $\matr{h}^{\rm t}$ outside the domain of \mbox{arccoth} [cf. Eq.~\eqref{eq:Hpar}] thereby precluding the correspondence to a parent Hamiltonian, and thus parent density matrix $\hat{\rho}=Z^{-1}\exp\left[-\beta \sum_{i,j} \hat{a}_i^\dag (\matr{H}_\text{par})_{ij} \hat{a}_j\right]$.

Whether $|\eta^{\rm t}_k|<1$ arise depends both on the degree of truncation and on the temperature. For  a thermal system at high temperature (small  $\beta$) the eigenvalues  $\eta_k = \coth({\beta}\epsilon_k/2)$ of $\matr{h}_T$ are far from $[-1,1]$ if the $\epsilon_k$ spectrum  is bounded. One then expects $|\eta^{\rm t}_k|>1$ to continue to hold under any moderate truncation. However for the zero temperature system, represented by the flat-band Hamiltonian $\matr{h}$  with eigenvalues  $\pm1$, getting $|\eta^{\rm t}_k|<1$ for at least some eigenvalues is inevitable however small the truncation is.

To address the cases where $|\eta^{\rm t}_k|<1$, we modify the truncated link matrix $\matr{h}^{\rm t}$  into a  ``reconstructed" link matrix $\matr{h}^{\rm t,r}$ by a shift and rescaling
\begin{equation}
\matr{h}^{\rm t,r}=\alpha(\matr{h}^{\rm t} -\gamma \matr{1})\,.
\end{equation}
Here $\alpha$ and $\gamma$ are chosen to ensure that the eigenvalues of this truncated and reconstructed link matrix,  $\eta^{\rm t,r}_k=\alpha(\eta^{\rm t}_k-\gamma)$, satisfy $|\eta^{\rm t,r}_k|\geq1$ (see below).
As $\matr{h}^{\rm t,r}$ has a spectrum in the domain of arccoth,  it now corresponds to an effective parent Hamiltonian $\beta^{\rm t,r} {{\matr{H}^{\rm t,r}}}$ with eigenvalues $\beta^{\rm t,r}\epsilon^{\rm t,r}_k=2\,\mathrm{arccoth}\eta^{\rm t,r}_k$.
After this reconstruction we obtain the approximate Grassmann formula
\begin{equation}\label{reconstruct}
\langle\hat X\rangle\approx\frac{\int(\prod_{j=1}^n\mathrm d\theta_j\mathrm d\bar\theta_j)\tilde X(\boldsymbol\theta,\bar{\boldsymbol\theta})\mathrm e^{-\bar{\boldsymbol\theta}^\top\matr{h}^{\rm t,r}\boldsymbol\theta}}{\int(\prod_{j=1}^n\mathrm d\theta_j\mathrm d\bar\theta_j)\mathrm e^{-\bar{\boldsymbol\theta}^\top\matr{h}^{\rm t,r}\boldsymbol\theta}},
\end{equation}
which replaces $\matr{h}_T$ by $\matr{h}^{\rm t,r}$ in Eqs.~\eqref{Grassmann} and \eqref{norm}.

To specify $\alpha$ and $\gamma$, we impose two constraints,
one each on $\eta^{\rm t,r}_+$ and $\eta^{\rm t,r}_-$, the minimum positive  and the maximum negative eigenvalue of $\matr{h}^{\rm t,r}$, respectively. These, via arccoth, correspond to the largest (in magnitude) positive and negative eigenenergies, $\epsilon^{\rm t,r}_{+}$ and $\epsilon^{\rm t,r}_{-}$, respectively, of $\matr{H}^{\rm t,r}$.
To set $\eta^{\rm t,r}_\pm$, we choose
$\beta^{\rm t,r}\epsilon^{\rm t,r}_{\pm}=\beta\epsilon_{\pm}$, that is,

\begin{equation}\label{constraints}
\eta^{\rm t,r}_{\pm}=\coth\left(\frac{\beta\epsilon_{\pm}}{2}\right),
\end{equation}
where $\epsilon_{+}$ and $\epsilon_{-}$ are maximum and minimum eigenenergies of $\matr{H}$, respectively, and $\beta$ is the physical inverse temperature.
For zero temperature,
as $\beta\epsilon_\pm\rightarrow \pm \infty$, the constraints become  $\eta^{\rm t,r}_\pm=\pm1$.
We note that the corresponding infinite thermal eigenvalues $\beta^{\rm t,r}\epsilon^{\rm t,r}_{\pm}$ only indicate the corresponding states of the approximate nonzero temperature problem always being occupied or empty, but not a divergent approximation for the total energy $\langle \hat{H}\rangle$, which should be calculated via  Eq.~\eqref{reconstruct}.

We also note that instead of Eq.~\eqref{constraints}, one could also place the $\eta^{\rm t,r}_k$ spectrum into the domain of arccoth via a requirement on $\min(|\eta^{\rm t,r}_-|,|\eta^{\rm t,r}_+|)$ only. As this is just one constraint, the parameters $\alpha$, $\gamma$ may be chosen such that an additional requirement is satisfied, e.g., that the exact average particle number is maintained under the truncation reconstruction procedure:
$\langle\hat N\rangle\!=\!\sum_{k=1}^n\frac{1}{\mathrm e^{\beta\epsilon_k}+1}$ be equal to $\langle\hat{N}^{\rm t,r}\rangle\!=\!\sum_{k=1}^n\frac{1}{\mathrm e^{2\mathrm{arccoth}\eta^{\rm t,r}_k}+1}$.

With the truncation-reconstruction procedure described above, a remaining question is whether the parent Hamiltonian $\matr{H}^{\rm t,r}$ is gapped. The gap of $\matr{H}^{\rm t,r}$ closes if the spectrum of $\matr{h}^{\rm t,r}$ approaches $\pm\infty$. Provided the spectrum of $\matr{h}_T$ is bounded (as is the case for insulators), no gap closing of $\matr{H}^{\rm t,r}$ will occur as the changes due to truncation (and thus the corresponding adjustments in reconstruction) are exponentially controlled.  We expect the gapless case to arise only near topological phase transitions, where the gap of the physical Hamiltonian $\matr{H}$ vanishes.

\begin{figure*}
\includegraphics[width=1.85\columnwidth]{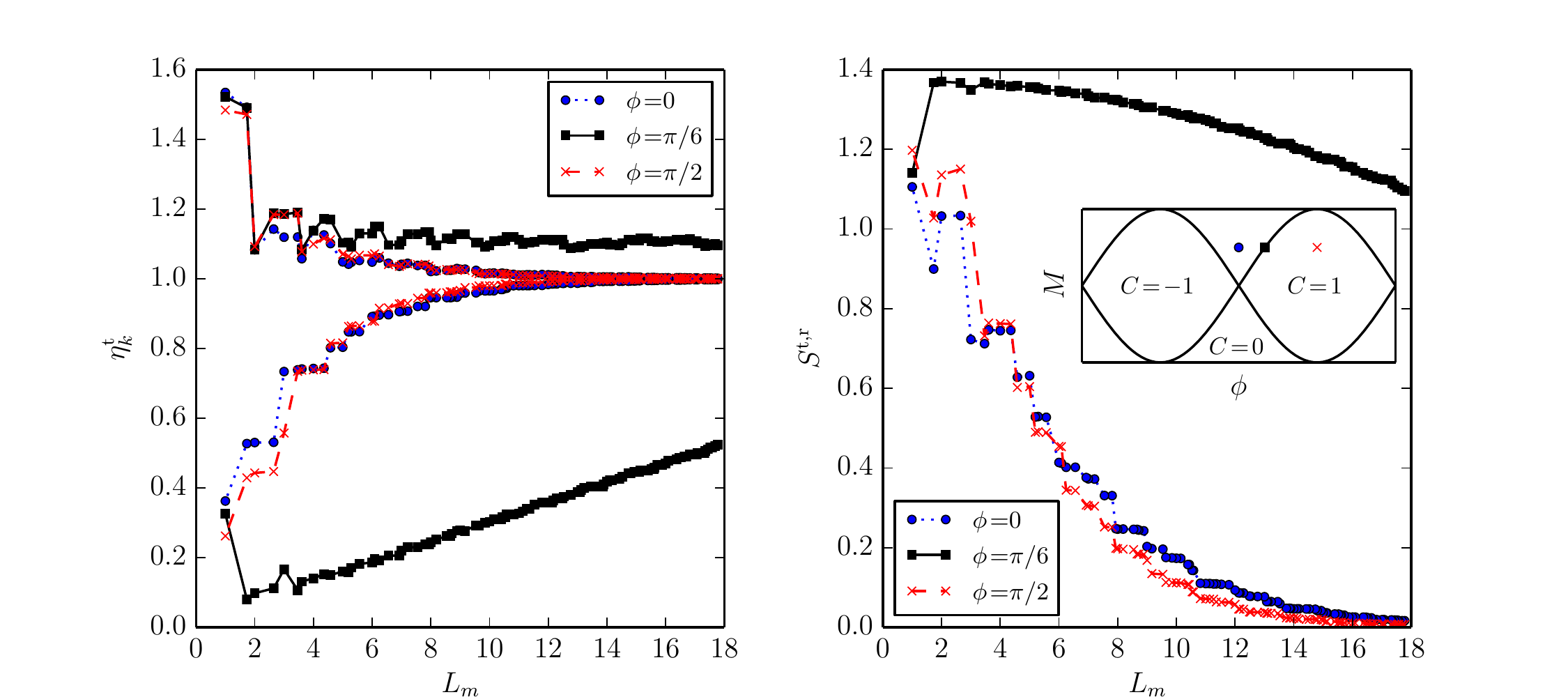}
\caption{Left: The smallest and largest positive eigenvalues $\eta^{\rm t}_k$ versus truncation length $L_m$
(in units of nearest neighbor distance)
for Chern  ($C=1$, crosses) and trivial ($C=0$, circles) insulating cases, and the transition between them (squares). The negative eigenvalues follow by spectral symmetry. The parameters are $t_1=1$, $t_2=1/3\sqrt{3}$, $M=0.5$, and $N=40$.
The lines are guides for the eye.
Right: The entropy $S^{\rm t,r}$ (per unit cell) as a function of $L_m$,
with the same parameters. }\label{spectrum}
\end{figure*}

\section{Numerical results for the Haldane model}
\label{sec:numerics}

We now turn to studying our truncation reconstruction procedure on a concrete model.
The 2D Haldane model on honeycomb lattice is a minimal, two-band, system permitting nontrivial topological phases \cite{Haldane:1988}. Its Hamiltonian involves on-site energy offsets $\pm M$ for A and B sublattices, nearest neighbour hoppings with uniform amplitude $t_1$, and next-nearest neighbour hoppings with coefficients $t_2\mathrm e^{\pm\mathrm i\phi}$ where the positive (negative) sign corresponds to the hoppings forming a closed triangle in one hexagon with clockwise (counterclockwise) direction. This model is known to yield Chern numbers $C=\pm1$ when $|M|<3\sqrt{3}|t_2\sin\phi|$ (cf. insets in Fig.~\ref{spectrum}). We utilize this model to demonstrate the truncation and reconstruction approach described above for gapped systems, including its applicability to topological phases. For comparison, we also show results for a gapless system arising at a topological phase transition.

In what follows we focus on the approximation of groundstate expectation values, i.e., the truncation starting from $\matr{h}$, where reconstruction is always necessary as discussed before. For simplicity, we choose a lattice of rhombus shape with periodic boundary conditions.

We can obtain the groundstate expectation value of any operator using Eq.~\eqref{Grassmann} with $\matr{h}_T$ replaced by the flat-band link matrix $\matr{h}$. As $\matr{h}$ is only exponentially but not strictly local,  we wish to truncate it into $\matr{h}^{\rm t} _m$ with various truncation lengths $L_m$ by setting all entries $h_{ij}\!=\!0$
if \mbox{$R_{ij}\!>\!L_m$.}
(We choose $L_m$ as the $m$-th nearest neighbor distance.)
Then we reconstruct to $\matr{h}^{\rm t,r}_m$, which is not only gapped around $(-1,1)$ by construction, but, as expected and verified numerically, also has finite bandwidth if $\matr{H}$ is gapped, thus corresponding to a gapped $\matr{H}^{\rm t,r}$.

\begin{figure*}
\includegraphics[width=1.95\columnwidth]{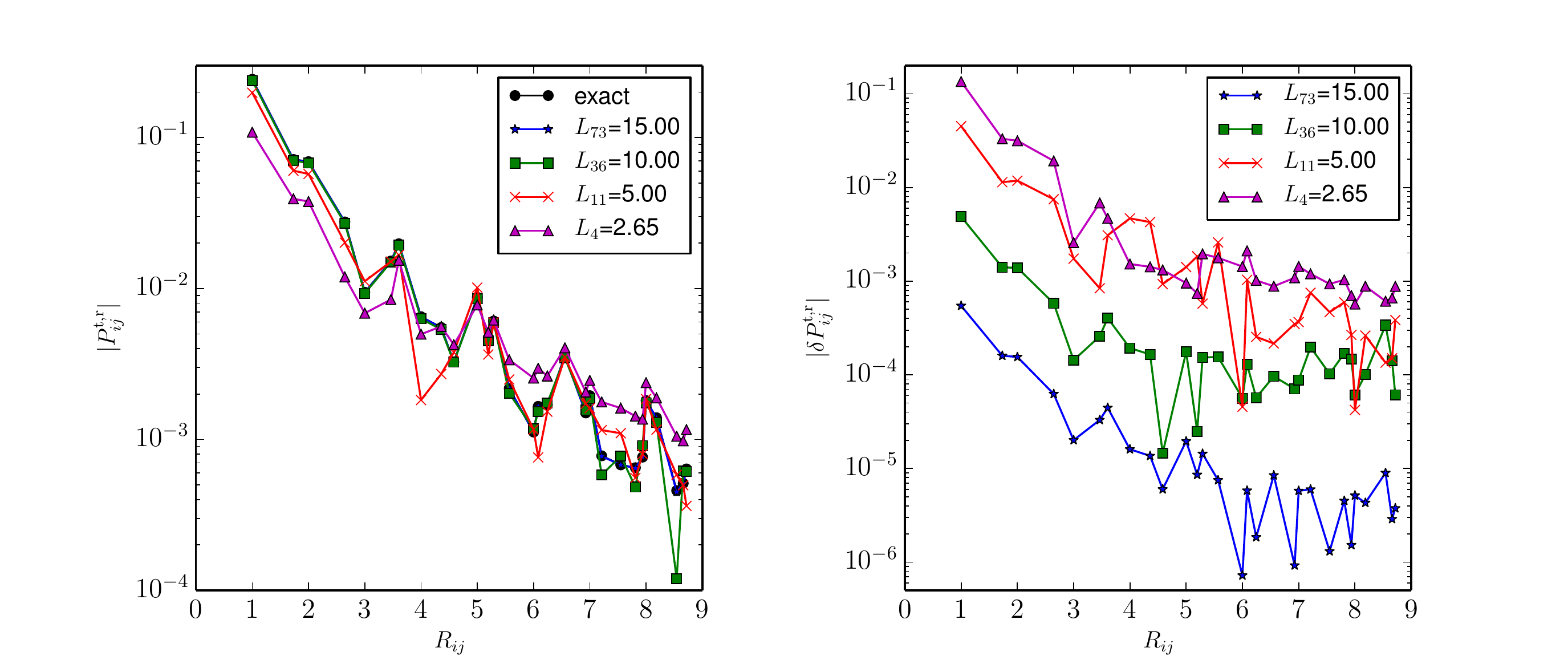}
\caption{Left: The absolute value of approximate two-point correlations $P^{\rm t,r}_{ij}$
versus distance $R_{ij}$
(in units of nearest neighbor distance)
for various truncation lengths $L_m$ as well as the exact result $P_{ij}$, for the Chern insulator  case of Fig.~\ref{spectrum} ($M=0.5$, $\phi=\pi/2$,  $C=1$). Right: the absolute value of the corresponding errors, $\delta P^{\rm t,r}_{ij}=P_{ij}-P^{\rm t,r}_{ij}$. The lines are guides for the eye.}\label{correlation}
\end{figure*}

For the simplicity of reconstruction, we consider the system at half filling; here $\matr{H}$ is particle-hole symmetric. The truncation procedure is chosen such that the same particle-hole symmetry holds for $\matr{h}^\text{t}$; this implies $\gamma\!=\!0$.  At half filling, this two-band model has average particle number $N_\mathrm{tot}$
for a system with $N_\mathrm{tot}=N\times N$ unit cells, where $N$ is the period along one edge of the rhombus; the preservation of spectral symmetry during truncation ensures that \mbox{$\langle\hat{N}^{\rm t,r}\rangle=N_\mathrm{tot}$} is maintained. We find that for gapped systems, the finite correlation length renders the $N$ dependence in our numerics negligible beyond a certain system size; the results
in Figs.~\ref{spectrum} and \ref{correlation} are for $N=40$ which is already in this $N$ independent regime.

We consider a number of measures to assess the effects of truncation. Since the spread and the location of the $\eta^{\rm t}_k$ spectrum  determines the properties of the subsequent reconstruction, in Fig.~\ref{spectrum} (left panel) we show the smallest and largest positive $\eta^{\rm t}_k$  as a function of truncation length $L_m$. (The behavior of the negative eigenvalues follows by spectral symmetry.) The rate of $\eta^{\rm t}_k$ deviating from $\pm1$ as $L_m$ is reduced depends on how far the system is from the topological phase transition. The closer to the transition, the wider the $\eta^{\rm t}_k$ spectrum becomes, as truncations result in larger errors due to  $\matr{h}$ becoming less local.
Due to $\gamma\!=\!0$,
and spectral symmetry,
the reconstruction amounts to $\eta^{\rm t,r}_k= \alpha\eta^{\rm t}_k$ with $\alpha=1/\eta^{\rm t}_+$. The $\eta^{\rm t,r}_k$ spectrum is thus a further factor of $1/\eta^{\rm t}_+$ wider than that of $\eta^{\rm t}_k$.

Next we turn to the entropy corresponding to $\beta^{\rm t,r} \matr{H}^{\rm t,r}$, which provides a useful single-number measure of how thermal the state corresponding to $\beta^{\rm t,r} \matr{H}^{\rm t,r}$ is and thus how far it is from the ground state of the physical Hamiltonian. We have
\begin{equation}
S^{\rm t,r}\!=\!-\sum_k[p^{\rm t,r}_k\ln p^{\rm t,r}_k+(1\!-\!p^{\rm t,r}_k)\ln(1\!-\! p^{\rm t,r}_k)],
\end{equation}
where %
$p^{\rm t,r}_k=[\exp(\beta^{\rm t,r}\epsilon^{\rm t,r}_k)+1]^{-1}$
and the summation is over all eigenstates.
The results for $S^{\rm t,r}$ (per unit cell) are shown in  Fig.~\ref{spectrum} (right panel). Similarly to the spectrum plots, the truncation effect is smaller for cases deep in the gapped (topological or nontopological) phases but larger near a phase transition. It is noticeable that $S^{\rm t,r}$ grows from $0$ with decreasing $L_m$ faster than the spectra do in Fig.~\ref{spectrum}, because the entropy formula increases dramatically from zero with $\eta^{\rm t,r}_k$ moving away from $\pm1$. In the extreme case at the topological phase transition, e.g., with $M\!=\!0.5$, $\phi\!=\!\pi/6$, $S^{\rm t,r}$ approaches the maximum for such a two-site unit cell, i.e., \mbox{$2\ln2\!\approx\!1.386$.}

Finally, to obtain a spatially resolved measure of the effects of truncation, we turn to the one-body density matrix $\langle\hat a_i^\dagger\hat a_j\rangle$, focusing on the gapped case where our approach is expected to work. Before truncation, as we are considering the ground state, \mbox{$\langle\hat a_i^\dagger\hat a_j\rangle=P_{ji}\!=\!h_{ji}/2$ ($i\!\ne\!j$).} Calculating the same after truncation and reconstruction using Eq.~\eqref{reconstruct} leads to the approximate result  \mbox{$\langle\hat a_i^\dagger\hat a_j\rangle^{\rm t,r}\!\equiv\!P^{\rm t,r}_{ji}=\![(\matr{h}^{\rm t,r})^{-1}]_{ji}/2$ ($i\!\ne\!j$).}  Given that $\gamma\!=\!0$ in the particular case we are studying, the change in the off-diagonal correlations after reconstruction is a simple scaling: $P^{\rm t,r}_{i\ne j}\!=\!\alpha^{-1}[(\matr{h}^{\rm t})^{-1}]_{i\neq j}/2$. In Fig.~\ref{correlation}, we show the approximate correlations $P^{\rm t,r}_{ij}$ and the corresponding errors, for the topologically nontrivial case of Fig.~\ref{spectrum} with $M=0.5$, $\phi=\pi/2$,  and Chern number $C=1$. (The topologically trivial insulator at $\phi=0$ has similar behavior.)
For both $P^{\rm t,r}_{ij}$ and the errors, the absolute values shown are isotropic, i.e., the spatial dependence is only through %
$R_{ij}$.
As the $L_m$ dependence of the errors shown illustrates, the approximation converges uniformly and exponentially upon increasing the truncation length;
the short-distance correlations are in the exponentially converging regime already for moderate $L_m$.
The accurate recovery of the one-body density matrix confirms that the Chern number $C$ is preserved throughout the truncation-reconstruction: owing to the exponential decay of correlations,  $C$ is encoded already in a few, short-range elements $P_{ij}$ \cite{Kitaev:2006,Ringel:2011} (a conclusion that remains true for thermal states as characterized in App.~\ref{app:rhotop} and references therein); this also allows the calculation of the Chern number via local Chern markers \cite{Bianco:2011,Caio:2019}. 
The preservation of $C$ is also as expected from the absence of gap closing during our procedure.

\section{Summary and Discussion}
\label{sec:summary}

We have formulated the calculation of thermal expectation values in free fermion systems in a form equivalent to Grassmann tensor networks, thereby generalizing the results of Ref.~\onlinecite{Beri:2011} to nonzero temperatures. The consideration of nonzero temperature averages allowed us to demonstrate how strictly local tensor networks can describe gapped, exponentially short-range parent Hamiltonians. Conversely, a tensor network for thermal averages was shown to be exponentially local provided we work with a gapped, strictly short-range Hamiltonian. Topology presents no obstructions to these results; they only require the corresponding spectrum to lie in the analytic domain of the function mapping between the link matrix and the parent Hamiltonian Eq.~\eqref{eq:Hpar} or vice versa, Eq.~\eqref{hH}. The parent Hamiltonians here, however, are Hamiltonians with respect to which nonzero temperature expectation values are calculated using the tensor networks, thus are understood in a slightly different sense than parent Hamiltonians in earlier works \cite{Wahl:2013,Dubail:2015,Read:2017} where tensor networks encode the corresponding ground states.

We have also described a truncation and reconstruction procedure applicable to the exact, exponentially local, tensor networks above, be those for thermal or groundstate averages. The strictly local tensor networks arising in this process approximate  exact averages by averages with respect to a sequence of parent Hamiltonians at fictitious nonzero temperatures; the corresponding thermal states approach their exact form upon increasing the truncation length.
We have illustrated this procedure by approximating groundstate averages for the 2D Haldane honeycomb model and found that the truncation-reconstruction works well both for nontopological and topological gapped phases.

While we formulated our analysis in terms of thermal states of Hamiltonians, our approach, at its core, only uses the Gaussian nature of the state of interest and therefore may be extended (by suitable choices for $f$ in Theorem~\ref{th:1}) to include Gaussian states arising, e.g., from dissipative dynamics~\cite{rhotop1,rhotop2}. In that context, a trade-off analogous to that observed here, namely between strictly short-range dissipative dynamics, nontrivial topology, and the purity of the achievable states, has been pointed out recently~\cite{Goldstein:2018}; the reason there is the same as for tensor-networks (cf. Sec.~\ref{sec:theoremHpar}): the impossibility for the flat band Hamiltonian $\matr{h}$ to be both strictly short-range and topologically nontrivial \cite{Chen:2013,Read:2017}. 

Our results represent the first step towards developing efficient approaches bridging tensor networks, mixed Gaussian states, and topology. Key to our approach is exploiting the advantages offered by focusing on expectation values directly and bypassing representations for states themselves. This initial study based on simple free-fermion Hamiltonian leaves much to be investigated in the future. One of the most interesting questions is to what extent our approach can be generalized to capture and approximate strongly correlated states.

\section*{Acknowledgements}

This research was supported by the NSFC grant No.~11704072,  the EPSRC grants EP/M02444X/1,  EP/K030094/1, EP/P034616/1 and EP/P009565/1, a Simons Investigator award, the Royal Society, and the ERC Starting Grant No.~678795 TopInSy.

\appendix

\section{Thermal expectation values via Grassmann integrals}

\label{app:grassmann}

In this Appendix we verify the Grassmann integral expression Eq.~\eqref{Grassmann}. The averages on both sides of Eq.~\eqref{Grassmann} can be calculated using Wick's theorem. %
The substitution rules
$\hat a_j\rightarrow\frac{\theta_j}{\sqrt{2}}$, $\hat a_j^\dagger\rightarrow\frac{\bar\theta_j}{\sqrt{2}}$, and $\hat a_j^\dagger\hat a_j\rightarrow\frac{\bar\theta_j\theta_j+1}{2}$ from the operator $\hat{X}(\hat a,\hat a^\dagger)$ to $\tilde X(\theta,\bar{\theta})$ thus imply the following correspondence between two-point averages:
\begin{equation}\label{eq:2corr}
\langle\hat a_j^\dagger\hat a_l\rangle=\frac{1}{2}\langle\delta_{jl}+\bar\theta_j\theta_l\rangle.
\end{equation}
Eq.~\eqref{Grassmann} holds if calculating the left hand side of Eq.~\eqref{eq:2corr} with the thermal density matrix gives the same result as calculating the right hand side via the Grassmann integral expression.

For the right hand side, the Grassmann integral yields
\begin{equation}\label{PGrassmann}
\frac{1}{2}\langle\delta_{jl}+\bar\theta_j\theta_l\rangle =\frac{1}{2}(\delta_{jl}-(\matr{h}_T^{-1})_{lj}) \,.
\end{equation}
To evaluate the left hand side, we define
\begin{equation}
C_{lj}=\langle\hat{a}_{j}^{\dagger}\hat{a}_{l}\rangle=\frac{1}{Z}\text{Tr}\left(\hat{a}_{j}^{\dagger}\hat{a}_{l}e^{-\beta\hat{H}}\right).
\end{equation}
In terms of the matrix $\mathsf{U}$ that diagonalizes the single-particle
Hamiltonian as $\mathsf{H}=\mathsf{U}^{\dagger}\mathsf{E}\mathsf{U}$ with $\matr{E}=\text{diag}(\epsilon_{k})$,
we have
\begin{equation}
\mathsf{C}=\mathsf{U}^{\dagger}\mathsf{D}\mathsf{U},
\end{equation}
with
\begin{equation}
\mathsf{D}_{kq}=\langle\hat{a}_{q}^{\dagger}\hat{a}_{k}\rangle=\delta_{kq}\langle\hat{a}_{k}^{\dagger}\hat{a}_{k}\rangle
\end{equation}
in terms of the annihilation (creation) operators $\hat{a}_{k}^{(\dagger)}$
of single-particle eigenstates. Using
\begin{equation}
\langle\hat{a}_{k}^{\dagger}\hat{a}_{k}\rangle=\frac{1}{\exp(\beta\epsilon_{k})+1}=\frac{1}{2}\left[1-\tanh\left(\frac{\beta\epsilon_{k}}{2}\right)\right],
\end{equation}
we find
\begin{equation}\label{eq:corrmatrix1}
\mathsf{C}=\frac{1}{2}\left[\mathsf{I}-\mathsf{U}^{\dagger}\tanh\left(\frac{\beta\matr{E}}{2}\right)\mathsf{U}\right]=\frac{1}{2}\left(\mathsf{I}-\mathsf{h}_{T}^{-1}\right),
\end{equation}
or
\begin{equation}\label{eq:corrmatrix2}
\langle\hat{a}_{j}^{\dagger}\hat{a}_{l}\rangle=\frac{1}{2}\left(\delta_{jl}-(\mathsf{h}_{T}^{-1})_{lj}\right)
\end{equation}
as required.

\section{Topological characterization of free fermion thermal states}
\label{app:rhotop}

In this Appendix, we briefly summarize how the topological features of free fermion thermal states may be characterized, adopting the approach of Ref.~\onlinecite{rhotop2} to the particle number conserving case we consider here. The topological characterization of Ref.~\onlinecite{rhotop2} is in terms of the correlation matrix $\Gamma_{ij}^{pq}=(i/2)\langle[\hat{c}_{i}^{p},\hat c_{j}^{q}]\rangle$, where $i,j=1,\ldots,n$, $p,q=1,2$ with Majorana fermions  $\hat c_{j}^{1}=\hat a_{j}+\hat a_{j}^{\dagger}$, $\hat c_{j}^{2}=(-i)(\hat a_{j}-\hat a_{j}^{\dagger})$. Collecting the $p,q$ components of $\Gamma_{ij}^{pq}$ in a matrix,
\begin{equation}\label{eq:Gammamatrix}
\Gamma_{ij}=i\left\langle \left(\begin{array}{c}
\hat c_{i}^{1}\\
\hat c_{i}^{2}
\end{array}\right)\left(\begin{array}{cc}
\hat c_{j}^{1} & \hat c_{j}^{2}\end{array}\right)\right\rangle -i\mathsf{I}_2\delta_{ij}.
\end{equation}
For a particle number conserving system, one can relate the correlation matrix to the correlator $\langle\hat{a}_{j}^{\dagger}\hat{a}_{l}\rangle$ using  \mbox{$W=\frac{1}{\sqrt{2}}\left(\begin{smallmatrix}
1 & 1\\
-i & i
\end{smallmatrix}\right)$} as
\begin{equation}
\!\!\!\!-iW^{\dagger}\Gamma_{ij}W\!=\left(\begin{array}{cc}
\delta_{ij}-2\langle \hat a_{j}^{\dagger}\hat a_{i}\rangle\\
 & 2\langle \hat a_{i}^{\dagger}\hat a_{j}\rangle-\delta_{ij}
\end{array}\right).
\end{equation}
For a free-fermion thermal state at temperature $\beta^{-1}$ and with single-particle Hamiltonian $\matr{H}$ this takes the form
\begin{equation}
-iW^{\dagger}\matr{\Gamma}W=\!=\!
\left(\begin{array}{cc}
\tanh\left(\frac{\beta \matr{H}}{2}\right)\\
 & -\tanh\left(\frac{\beta \matr{H}^T}{2}\right)
\end{array}\right).
\end{equation}
Viewing $\matr{\Gamma}$ as a single-particle Hamiltonian, it is gapped around zero energy if and only if $\matr{H}$ is an insulator. For such gapped $\matr{\Gamma}$, the topological properties (determined by invariants whose concrete form depends on the presence or absence of time-reversal and sublattice symmetries as well as the dimensionality of the system) are set by the ``ground state" (negative ``energy") \mbox{eigenvectors of $\matr{\Gamma}$ \cite{rhotop2}.} Recognising the right-hand side as the Bogoliubov-de-Gennes Hamiltonian corresponding to $\tanh(\beta\matr{H}/2)$, we note that, up to a particle-hole redundancy, these eigenvectors have the same topology as the groundstate eigenvectors of $\tanh(\beta\matr{H}/2)$. These, in turn, are the same as those of $\matr{H}$ due to $\tanh(x)$ being a monotonically increasing odd function of $x$. Thus, the topology of the groundstate eigenvectors of $\matr{H}$ also determines that of the thermal density matrix.

\section{Relation to Grassmann tensor networks}
\label{app:GTNrel}

\begin{figure}
\includegraphics[width=0.85\columnwidth]{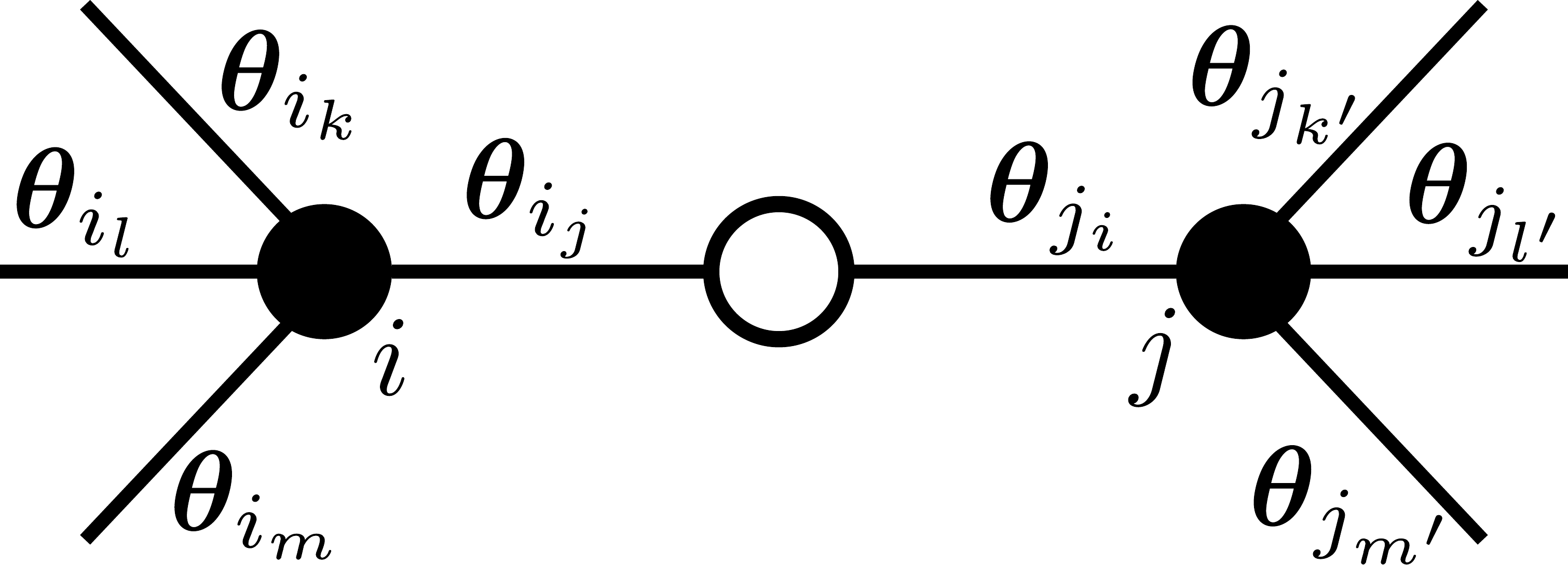}
\caption{The structure of Grassmann tensor networks for expectation values~\cite{Gu:2010}. Full circles denote the site tensors $T^{(X_i)}_i$ and $T^{(X_j)}_j$; the empty circle denotes the link tensor $G_{ij}$. The Grassmann variables $\boldsymbol{\theta}_J=\{\theta_J, \bar\theta_J\}$ of the link tensors (and the corresponding differentials of the site tensors) reside on the legs $J$ of the sites.}\label{fig:TN}
\end{figure}

In this Appendix, we briefly review the relation between our expressions and contractions of the Grassmann tensor networks introduced in Ref.~\onlinecite{Gu:2010}. The structure of such network contractions, specialized for the averages we consider, is illustrated in Fig.~\ref{fig:TN}: links $ij$ between sites $i$ and $j$ are associated with link tensors $G_{ij}$; sites are associated with site tensors $T_{j}^{(X_j)}$
(the superscript indicates that the concrete form of these depends on the action $X_j$ of operator $X$ on site $j$). The Grassmann variables $\theta_{j_{i}}$, $\bar{\theta}_{j_{i}}$ of the network are associated with the legs $j_{i}$ of sites $j$ (with $j_{i}$ denoting the end of link $ij$ running into site $j$). The link tensors $G_{ij}$ are Grassmann even expressions of $\theta_{i_{j}},\bar{\theta}_{i_{j}},\theta_{j_{i}},\bar{\theta}_{j_{i}}$; the site tensors $T_{j}^{(X_j)}$ are expressions involving Grassmann differentials $\{d\theta_{J},d\bar{\theta}_{J}|J\in j\}$ with $J\in j$ denoting leg $J$ of site $j$. The Grassmann parity of $T_{j}^{(X_j)}$ equals the fermion parity of $X_j$.

The tensor network contractions are of the form 
\begin{equation}
\langle X(\hat{a},\hat{a}^{\dagger})\rangle=N\,P_{0}\int\prod_{j}T_{j}^{(X_j)}\prod_{k<l}G_{kl},
\end{equation}
where $P_{0}$ denotes a projection on the scalar part of the resultant
integral and where we separated a normalization factor as in Eq.~\eqref{Grassmann}. As described in Ref.~\onlinecite{Gu:2010}, such contractions may be efficiently approximated using the Grassmann tensor entanglement renormalization group; for works applying this scheme to concrete problems see e.g., Refs.~\onlinecite{Gu:2013a},~\onlinecite{Gu:2013b}.   

The conversion of the expression Eq.~\eqref{Grassmann} or Eq.~\eqref{GTN-tostart} is achieved using a combination of the Grassmann Leibniz rule \cite{Zinn-Justin:2002} and inserting suitable scalar part projections; this transforms the initial integral from one over Grassmann variables on the sites of the lattice into one over an increased number of Grassmann variables on the legs, as in Fig.~\ref{fig:TN}. This leads to link tensors [in terms of Eq.~\eqref{GTN-tostart}, for concreteness]
\begin{equation}
G_{kl}=\exp[-\bar{\theta}_{k_{l}}\theta_{l_{k}}g_{kl}]\exp[-\bar{\theta}_{l_{k}}\theta_{k_{l}}g_{lk}].
\end{equation}
The site tensors  where $X$ does not act ($X_j=1$) are 
\begin{equation}
T_{j}^{(1)}=\left(\sum_{J\in j}d\theta_{J}\right)\left(\sum_{J\in j}d\bar{\theta}_{J}\right)-g_{jj}.
\end{equation}
For $X_j=\hat{a}_{j}$, $X_j=\hat{a}_{j}^{\dagger}$
, and $X_j=\hat{a}_{j}^{\dagger}\hat{a}_{j}$ we have
\begin{equation}
T_{j}^{(\hat{a}_{j})}=-\frac{1}{\sqrt{2}}\sum_{J\in j}d\bar{\theta}_{J},\quad T_{j}^{(\hat{a}_{j}^{\dagger})}=\frac{1}{\sqrt{2}}\sum_{J\in j}d\theta_{J},
\end{equation}
and
\begin{equation}
T_{j}^{(\hat{a}_{j}^{\dagger}\hat{a}_{j})}=\frac{1}{2}\left[\left(\sum_{J\in j}d\theta_{J}\right)\left(\sum_{J\in j}d\bar{\theta}_{J}\right)-(g_{jj}-1)\right],
\end{equation}
respectively. We emphasize, however, that for our purposes the conversion to the structure in Fig.~\ref{fig:TN} is not required; we directly work with the integrals of the form  Eq.~\eqref{GTN-tostart} and the resultant matrix expressions. 

\section{Proof of Theorem \ref{th:1}}

\label{app:bcb}

In this Appendix we briefly outline the proof of Theorem~\ref{th:1}. This is based on the proof of Theorem 8.12 of Benzi {\it et al.} \cite{Benzi:2012} and its matrix generalizations discussed in the same reference.

A preliminary step is to show that a function $\tilde{f}$ that takes real values on $[\eta_1^2,\eta_2^2]$, and is analytic in a neighbourhood of $[\eta_1^2,\eta_2^2]$ in the complex plane, can be approximated by polynomials to exponential accuracy.
Based on this, one can show that $f$ in Theorem~\ref{th:1} can also be approximated by polynomials to exponential accuracy. Then, the matrix decay bound in Theorem~\ref{th:1} follows via the spectral decomposition of $\matr{M}$.
Our proof only differs from that of Benzi {\it et al.} \cite{Benzi:2012} in the preliminary step,
and only in the choice of the ellipse $\mathcal{E}_\chi$ below, which accommodates the slightly weaker analyticity requirements on $\tilde{f}$ (and thus on $f$) compared to those in Ref.~\onlinecite{Benzi:2012}.

To make our formulation more precise, for integer \mbox{$k\geq0$,} we define $E_k(\varphi,I)$ as the error for the $k$-th best approximation of a function $\varphi$ on a finite set $I$ of finite intervals:
\begin{equation}
E_k(\varphi,I)=\inf\left\{\max_{x\in I}|\varphi(x)-p(x)|:p\in P_k\right\},
\end{equation}
where $P_k$ is the set of all polynomials with real coefficients of degree no greater than $k$.

Consider now an ellipse $\mathcal{E}_\chi$ with foci at $\eta_1^2$ and $\eta_2^2$, and denote its semiaxes by $a$ and $b$ ($a\!>\!b$). With the constraint $a^2\!-\!b^2\!=\!c^2$ where the fixed semifocal length $c\!=\!\frac{\eta_2^2\!-\!\eta_1^2}{2}$, the ellipse is parametrised by $\chi\!=\!\frac{a\!+\!b}{c}$, which determines $a\!=\!\frac{\eta_2^2-\eta_1^2}{4}(\chi\!+\!\chi^{-1})$ and \mbox{$b\!=\!\frac{\eta_2^2\!-\!\eta_1^2}{4}(\chi\!-\!\chi^{-1})$.} The infimum of $\chi$ is $1$, when $\mathcal{E}_\chi$ reduces to the segment $[\eta_1^2,\eta_2^2]$ on the real axis. The supremum $\bar\chi$ is set by the analytic properties of $\tilde{f}$; it corresponds to $\mathcal{E}_\chi$ first touching a singularity of $\tilde{f}$ as it is gradually blown up with increasing $\chi$. By the coordinate transformation
\begin{equation}
x(u)=\frac{\eta_2^2\!-\!\eta_1^2}{2}u+\frac{\!\eta_1^2\!+\!\eta_2^2}{2}
\end{equation}
we map $u\in [-1,1]$ to $x\in [\eta_1^2,\eta_2^2]$ and the ellipse $\mathcal{E}^0_\chi$ with vertices  $\pm(\chi\!+\!\chi^{-1})/2$ and $\pm\mathrm i(\chi\!-\!\chi^{-1})/2$ to $\mathcal{E}_\chi$. \mbox{For $1<\chi<\bar{\chi}$,} the function $\tilde{f}[x(u)]$ is analytic on $\mathcal{E}^0_\chi$ and takes real values on $[-1,1]$. One can thus invoke Bernstein's theorem \cite{Bernstein} (see also Theorem 8.7 in Ref.~\onlinecite{Benzi:2012}):
\begin{equation}
E_k(\tilde{f}[x(u)],[-1,1])\leq \frac{2M(\chi)}{\chi^k(\chi-1)},
\end{equation}
where $M(\chi)=\max_{z\in \mathcal{E}_\chi}|f(z)|$.
Because $x(u)$ is a linear transformation, we also have
\begin{equation}
E_k(\tilde{f}(x),[\eta_1^2,\eta_2^2])\leq \frac{2M(\chi)}{\chi^k(\chi-1)}.
\end{equation}

Now taking $\chi$ such that $\pm\sqrt{\mathcal{E}_\chi}$ are in the interiors of the neighborhoods in Theorem~\ref{th:1}, the proof regarding the polynomial approximation of $f$ on $D$ is identical to that of Theorem 8.12 in Ref.~\onlinecite{Benzi:2012}: this uses suitable combinations of $f_{1(2)}$ and $\sqrt{\cdot}$ for $\tilde{f}$ to show that there exists $K>0$ such that
\begin{equation}\label{eq:Benzif}
E_k(f,D)\leq K\chi^{-k/2}.
\end{equation}

The final step is to use Eq.~\eqref{eq:Benzif} to prove Theorem~\ref{th:1}. As the strictly short-range Hermitian matrix $\matr M$ has range $R^*$, the matrix power $\matr{M}^k$ has range $kR^*$, and
so does the polynomial function $p_k(\matr{M})$ with $p_k(x)$ the $k$-th best approximation of $f(x)$ on $D$. Now consider a pair of sites $i\neq j$, and take $k$ to be the largest integer such that
$(k+1)R^*\geq R_{ij}> kR^*$.
The matrix elements $[p_k(\matr{M})]_{ij}$ are zero. Therefore,
\begin{equation}\label{eq:matrnorm}
|f(\matr{M})_{ij}|=|[f(\matr{M})-p_k(\matr{M})]_{ij}|\le||f(\matr{M})-p_k(\matr{M})||_2,
\end{equation}
where we use the matrix spectral norm,
\mbox{$||\matr{A}||_2=\max_{||x||=1}\sqrt{\langle x|\matr{A}^\dagger \matr{A}|x\rangle}$} or, equivalently, the maximum singular value of $\matr{A}$. (The inequality $|A_{ij}|\leq||\matr{A}||_2$ follows, e.g., from the Cauchy-Schwartz inequality applied to $A_{ij}=\langle e_i|\matr{A} e_j\rangle$ with basis vectors $|e_{i,j}\rangle.$)
As $\matr{M}$ is Hermitian, we can diagonalise it using a unitary matrix $\matr{U}$, i.e., $\matr{M}=\matr{U}^\dagger\Lambda\matr{U}$ where $\matr{\Lambda}$ is a diagonal matrix containing the eigenvalues $\lambda_j$. We also have $f(\matr{M})=\matr{U}^\dagger f(\Lambda)\matr{U}$ and $p_k(\matr{M})=\matr{U}^\dagger p_k(\Lambda)\matr{U}$.
As $f(\matr{M})-p_k(\matr{M})$ is also Hermitian, its singular values are the absolute values of its eigenvalues, $|f(\lambda_j)-p_k(\lambda_j)|$. Furthermore,
\begin{multline}||f(\matr{M})-p_k(\matr{M})||_2=\max_{\lambda_j}|f(\lambda_j)-p_k(\lambda_j)|\\ \leq \max_{x\in D}|f(x)-p_k(x)|\leq K\chi^{-k/2}\leq
K\chi^{-\frac{R_{ij}-R^*}{2R^*}},
\end{multline}
where the first inequality holds because $\lambda_j \in D$,
and
the second  because $p_k$ is the $k$-th best polynomial approximation.
Defining $\tilde{K}=K\chi^{1/2}$ and $\tilde{\chi}=\chi^{1/2}$ we find
\begin{equation}
|f(\matr{M})_{ij}|\leq \tilde{K}\tilde{\chi}^{-R_{ij}/R^*},
\end{equation}
which, upon renaming variables, proves Theorem~\ref{th:1}.

\end{document}